\documentclass[9pt,a4paper]{extarticle}

\newtheorem{condition}{\bf Condition}[section]

\usepackage{amssymb,bm} 
\usepackage{amsmath} 
\usepackage{amsthm} 

\numberwithin{equation}{section} 

\usepackage{graphicx}

\newcommand\commentout[1]{}

\newlength{\unitlengthstored}

\newcommand\say[1]{`{#1}'} 

\newcommand{\Rev}[1]{{                    #1}}

\newtheoremstyle{MyTheoremStyle}
{3pt}
{3pt}
{}
{}
{\bf}
{:}
{.5em}
{}
\theoremstyle{MyTheoremStyle}

\newtheorem{lemma}{Lemma}
\newtheorem{theorem}{Theorem}[section]

\newenvironment{remark}{}

\newcommand\myproof[1]{\vspace{3pt}\noindent{\bf Proof:} #1 \ \hfill  $ \blacksquare$\vspace{3pt}}

\newcommand{\Reynolds}{\ensuremath{Re}}
\newcommand\Rey{\Reynolds}  
\newcommand{\Nuss}{\operatorname{\mathit{N\kern-.2em u}}}

\newcommand\eq[1]{(\ref{#1})}

\newcommand\der[2]{\frac{d #1}{d #2}}
\newcommand\pd[2]{\frac{\partial #1}{\partial #2}}
\newcommand\pdd[2]{\frac{\partial^2 #1}{\partial {#2}^2}}

\newcommand\innerproduct[2]{\left\langle#1,#2\right\rangle}

\newcommand\varder[2]{\frac{\delta #1}{\delta #2}}
\newcommand\norm[1]{\left|\left|#1\right|\right|}

\newcommand\meanT[1]{\overline{#1}}
\newcommand\meanH[1]{\left\langle{#1}\right\rangle_h}
\newcommand\meanHT[1]{\overline{\meanH{#1}}}

\newcommand\zint[1]{\int_0^1{#1}\,dz}

\newcommand\ba{{\mathbf u}}
\newcommand\bA{\mathbf U}
\newcommand\br{{\mathbf r}}

\newcommand\bu{{\mathbf u}}

\newcommand\bvv{{\mathbf {\tilde  u}}}
\newcommand\tu{{ {\tilde  u}}}

\newcommand\tw{{ {\tilde  w}}}
\newcommand\bff{\mathbf f}

\newcommand\bl{\mathbf l}
\newcommand\bb{\mathbf b}

\newcommand\bU{\mathbf U}
\newcommand\bW{\mathbf W}
\newcommand\bD{{\mathbf D}}
\newcommand\bR{{\mathbf R}}

\newcommand\bxi{{\bm\xi}}
\newcommand\bomega{{\bm\omega}}
\newcommand\dissiIns{\epsilon}
\newcommand\dissiAve{{\overline\varepsilon}}
\newcommand\seis{s}

\newcommand\OTP{\overline{\text{Conv}}(P)}

\DeclareMathOperator*{\argmin}{arg\,min}
\DeclareMathOperator{\rot}{rot}

\begin{document}

\title{Relationship between the methods of bounding time averages}

\author{
S. Chernyshenko\\
Department of Aeronautics, Imperial College London, \\London SW7 2AZ, UK\\
s.chernyshenko$@$imperial.ac.uk}







\maketitle

\begin{abstract}
The problem of finding bounds of time-averaged characteristics of dynamical systems, such as the bound on the mean energy dissipation rate in a turbulent flow governed by incompressible Navier-Stokes equations, is considered. It is shown that the direct method described by Seis (\emph{J. Fluid Mech.}, {\bf 777}:591–603, 2015) and the  auxiliary functional method by Chernyshenko \emph{et al.}~(\emph{Phil. Trans. R. Soc. B.}, {\bf 372}:20130350, 2014) are related and can lead to the same bound. The well-known background flow method of Doering  and Constantin is equivalent to the auxiliary functional method with a quadratic auxiliary functional. The known implementations of the direct method apparently also correspond to quadratic auxiliary functionals. The findings are illustrated by the analysis of the plane Couette flow.
Three routes of further progress using non-quadratic auxiliary functionals and at the same time allowing to utilise the experience accumulated with the background flow method are proposed: making the balance parameter dependent on the energy, making the background flow time-dependent in a specific way, and adding helicity to the auxiliary functional. 

\end{abstract}

\section{Introduction}

One of the main attractions of the methods of bounding time averages is their promise of a rigorous solution of the problem of turbulence. This problem, the most widely known incarnation of which is associated with fluid flows at high Reynolds number, is encountered in dynamical systems so complicated that numerical calculations are too expensive even with the best computers available. A complete numerical solution, if obtained, would provide full information about the system behavior, while in practice it is often sufficient to have information about one or very few time-averaged quantities, such as the mean drag and lift.



\noindent The problem of turbulence is the problem of establishing methods of obtaining this limited information at a significantly smaller cost than needed for getting the complete solution. Thus, for a time-averaged quantity of interest $\meanT{F}$ finding the upper and lower bounds $B_L\le \meanT{F}\le B_U$ is a possible solution, in particular if there is a satisfactory trade-off between the margin $B_U-B_L$ and the computational cost. The history of the topic and its significance are covered extensively in the introduction of~\cite{Seis:JFM2015} and will not be repeated here, but it is worth mentioning that the method of~\cite{Seis:JFM2015}, as it is pointed out in this paper, stems from~\cite{otto2011rayleigh}. 
The present paper 
aims at revealing the connection between the widely-used background flow method~\cite{DC:PhysRev1994} and the newer methods proposed in~\cite{Seis:JFM2015} and~\cite{CGHP:RoySoc2014}
by recasting them in the framework of the method proposed in~\cite{CGHP:RoySoc2014}. This will allow us to suggest a few ways for further progress in this field.

\Rev{
We will use  the names \say{direct method} for the method of~\cite{Seis:JFM2015} and \say{auxiliary functional method} for the method
 of~\cite{CGHP:RoySoc2014}
\footnote{ 
\Rev{
 The direct method~\cite{Seis:JFM2015} was called  \say{another method} and \say{combined maximum
principle and background field method} in~\cite{nobili2017limitations}. The auxiliary functional method was not given a name in~\cite{CGHP:RoySoc2014} and was called  an \say{indefinite storage functional method} in~\cite{SC2017arXiv1704.02475}  (which is an early version of a part of the present paper).
}
}.
}
%
While the background flow method is well established, the other two approaches are new, and their exact definitions remain somewhat uncertain. 
This is particularly applicable to the direct method, and we suggest a more precise definition in this paper.

It has already been known that the background flow method is a special case of the auxiliary function method, but the relationship of these two methods 
 and the direct method is not obvious: in~\cite{Seis:JFM2015} it is stated: \say{Our new approach presented in this paper is entirely different from the background flow method}. 
To make the link between the methods as clear as possible, in Section~\ref{seq:general} we will aim at a maximum clarity at the expense of the problem-specific detail. To the same end we will not repeat the usual disclaimers concerning the existence and regularity: limits, functions, functionals, and solutions involved are silently assumed to exists where required and be regular enough for the operations performed.

Section~\ref{seq:general} contains the generic analysis of the relationships between the methods in question, including the demonstration of their effective equivalence. Precisely, we prove that any bound obtained by the auxiliary functional method can also be obtained by the direct method and that the converse is true under a certain condition. In that section we also reiterate that the background flow method is a particular case of the auxiliary functional method with the auxiliary functional of a special, `quadratic' as defined later, form.
The main statements in this section also apply in the finite-dimensional case with functionals and operators replaced with functions, variational derivatives replaced with gradients, and inner products replaced with dot products, so that one could write $V(\ba)$ instead of $V[\ba]$ and $\bff\cdot\nabla V$ instead of $\innerproduct{\bff}{\frac{\delta V}{\delta\ba}}$, which might make the text easier to understand. 
 Section~\ref{seq:PCF} illustrates the findings on the example of the plane Couette flow.
 Section~\ref{sec:discussion} proposes new ideas for further progress.

\section{The problem, the three methods and their relationship}\label{seq:general}


The problem addressed in the present work is formulated in the following way. 
Consider an infinite-dimensional dynamical system, the state of which is described by a vector $\ba$, which is an element of  an inﬁnite-dimensional function space. The dynamical system then determines the time derivative of $\ba$:

\begin{equation}\label{eqn:dyn_sys}
\der{\ba}t=\bff(\ba).
\end{equation}
  Let $F[\ba]$ be a functional the time-average of which is of interest. The problem is to find an upper bound $B$ such that $\meanT{F[\ba(t)]}\le B$ for any solution $\ba(t)\in A$ of (\ref{eqn:dyn_sys}), where the overline denotes the infinite time average. 

\begin{remark} 
Here, $\forall\ba\in A$ means that  for each time $t$,  $\ba(t)$  belongs to a set $A$ that is a~priori known to contain all solutions of~(\ref{eqn:dyn_sys}) we are interested in. For example, for fluid dynamical problems in which $\ba$ is the velocity field, $A$ can be assumed to be the set of all sufficiently regular incompressible velocity fields satisfying the correct boundary conditions.   
 Reducing the set $A$ can lead to better bounds. Reducing this set can be achieved by an independent analysis of (\ref{eqn:dyn_sys}), or by the researcher restricting the study to some but not all possible solutions.  In what follows we will often omit $\in A$ in $\forall \ba\in A,$ but it will always be implied.
\end{remark}

\subsection{Auxiliary functional method}\label{seq:StorageFunctionalMethod}

Let $V[\ba]$ be a differentiable functional of $\ba
$.  Then a functional $D[\ba]$ can be defined as the Lie derivative of $V[\ba]$ with respect to $\bff$, that is as
\[
D[\ba]=\innerproduct{\bff}{\frac{\delta V}{\delta\ba}}.
\]
 Here,  ${\delta V}/{\delta\ba}$ denotes the gradient of $V$ in the vector space of $\ba,$ and~$\innerproduct{\cdot}{\cdot}$ denotes an inner product in this space
. 

If the solution  $\ba(t)$ of (\ref{eqn:dyn_sys}) is substituted into $V$ then $V$ becomes a function of time, and its derivative can be calculated as $dV/dt=D[\ba(t)]$.  
 %
In problems involving time averages $\ba(t)$ are usually bounded, and, hence, so is $V[\ba(t)].$    
 \begin{lemma}\label{Lemma1} 
 For any bounded differentiable $V[\ba],$ if  
$\displaystyle
 D[\ba]=\innerproduct{\bff}{\frac{\delta V}{\delta\ba}}
$ 
  then
$\displaystyle
 \meanT{D[\ba(t)]}=0.
$ 
 \end{lemma} 
  \myproof{  
$\displaystyle
 \meanT{D[\ba(t)]}=\lim_{T\to\infty}\frac1T\int_0^T D[\ba(t)]\,dt=
 \lim_{T\to\infty}\frac{V[\ba(T)]-V[\ba(0)]}T=0.
$
}

 \begin{theorem}\label{MainTheorem}
If there exists a bounded differentiable $V[\ba]$ and a constant $B$ such that 
 \begin{equation}\label{eqn:constraint}
 F[\ba]+D[\ba]\le B\qquad \forall\, \ba\in A,
 \end{equation}
where $D[\ba]=\innerproduct{\bff}{\frac{\delta V}{\delta\ba}}$,
 then $\meanT{F[\ba(t)]}\le B$, that is $B$ is an upper bound for the time average of $F[\ba(t)].$
 \end{theorem}
\myproof{By Lemma~\ref{Lemma1}  $\meanT{D[\ba(t)]}=0$.}

The auxiliary functional\footnote{ 
If $V[\ba]\ge0$ for all $\ba,$  and $\ba$ is finite-dimensional then by definition (see~\cite{Willems1972})  $V[\ba]$  is a storage function with supply rate $B-F[\ba].$ 
For an auxiliary functional  the requirement  $V[\ba]\ge0$ is not needed. However, if  $V[\ba]$ proves a bound then under mild conditions $V[\ba]$  is  bounded from below, see A.2 in~\cite{Goluskin_2019}.  
}
 method~\cite{CGHP:RoySoc2014} of finding bounds for time averages consists in finding $B$ and $V[\ba]$ satisfying~\eq{eqn:constraint}. A lower bound can be obtained similarly.

Note that the corresponding  optimisation problem of finding the best possible upper bound,
 \begin{equation}\label{eqn:optimisation}
 B_\text{opt}=\min_{\substack{B,V\\ \text{s.t. \eq{eqn:constraint}}}}B,
  \end{equation} 
  is convex.

\subsection{Quadratic auxiliary functional}

In certain cases, such as that of incompressible
flows governed by the Navier–Stokes equations, the right-hand side of~\eq{eqn:dyn_sys}
 is a quadratic
nonlinear function that can be represented as
$
\bff(\ba)=\bb(\ba,\ba)+\bl(\ba)+\br
$, 
  where $\bb$ is a bilinear operator%
, $\bl$ is a linear operator and $\br$ is a constant term. One can look for the auxiliary functional in a similar form 
\begin{equation}\label{eqn:QuadraticV}
V[\ba]=Q[\ba,\ba]+L[\ba]+\text{const},
\end{equation}
where $Q[\ba,\ba]$ is a bilinear form, and $L[\ba]$ is a linear functional.   For functionals of the form~\eq{eqn:QuadraticV} $V(c\ba)$ is a quadratic polynomial of the scalar $c$. We will call such functionals quadratic. This terminology is establishing itself in particular in the works where the bound optimisation problem~\eq{eqn:optimisation} is solved using polynomial sum-of-squares optimisation~\cite{fantuzzi2021background}. 
Many quantities of interest, as for example the energy dissipation rate in fluid flows, are quadratic functionals.
\begin{theorem}\label{Theorem2}
If $\bff(\ba)$, $F[\ba]$ and $V[\ba]$ are quadratic
and satisfy \eq{eqn:constraint},  and the admissible set $A$ in \eq{eqn:constraint} is such that $c\ba\in A$ for any $\ba\in A$ and any scalar $c$  then 
 \begin{equation}\label{eqn:dEdt}
\innerproduct{\bb(\ba,\ba)}{\frac{\delta Q[\ba,\ba]}{\delta\ba}}=0\quad \forall{\ba}\in A.
\end{equation}
\end{theorem}
\myproof{  Since $V[\ba]$ is quadratic, $\delta V/\delta \ba$ is linear in $\ba.$ Hence,  $\innerproduct{\bff[c\ba]}{\delta V[c\ba]/\delta \ba}\sim c^3$ as $c\to\pm\infty$  and, therefore, \eq{eqn:constraint} cannot be satisfied in the general case~\cite{GC:PhysD2012}. The cubic terms in $\innerproduct{\bff}{\delta V/\delta \ba}$ can be identically zero if and only if (\ref{eqn:dEdt}) is satisfied. }

In simple words, Theorem~\ref{Theorem2} states that the nonlinear part of the dynamical system should conserve the homogeneous quadratic part of the quadratic auxiliary functional. This theorem can be extended by introducing higher-order polynomial auxiliary functionals, and it will then demand the nonlinear part of the system to conserve the highest-order homogeneous part of the functional.

Under the assumptions of Theorem~\ref{Theorem2}, 
\begin{equation}\label{eqn:DbySeis}
D[\ba]=\innerproduct{
\bl(\ba)+\br}{\varder{Q[\ba,\ba]}{\ba}}+\innerproduct{\bff(\ba)}{\varder{L(\ba)}{\ba}}.
\end{equation}
In fluid dynamics it is typical that the nonlinear part of the governing equations conserves the kinetic energy $\norm{\ba}^2/2=\innerproduct{\ba}{\ba}/2$, and then one can take
 $Q[\ba,\ba]=\text{const}\cdot
\norm{\ba}^2/2$. We will discuss whether this is the only possible choice in \S\ref{sec:discussion}\ref{sec:helicity} devoted to helicity.

\subsection{Background flow method and auxiliary functional method}
The background flow method~\cite{DC:PhysRev1994} for an incompressible fluid flow  
can be interpreted~\cite{CGHP:RoySoc2014} as an auxiliary functional method with the auxiliary functional of the form 
\begin{equation}\label{eqn:Vbackground}
V[\ba]=\alpha E[\ba-\bA]=\alpha\frac{\norm{\ba-\bA}^2}2
=\alpha \frac{\norm{\ba}^2-2\innerproduct{\ba}{\bA}+ \norm{\bA}^2}2,
\end{equation}
 where $\bA$ is a tunable parameter called the background flow, and $\alpha$ is a tunable constant called a balance parameter. The name is justified,
 since $\bA$ often belongs to the same admissible set $A$ as $\ba$ and stems from the particular applications in which $\ba$ is the velocity field of the flow in question.
By taking a variational derivative 
we obtain
\begin{equation}\label{eqn:D_byDC}
D[\ba]=
\alpha\innerproduct{\bl(\ba)+\br}{\ba}-\alpha\innerproduct{\bff(\ba)}{\bA}
\end{equation}
(see (\ref{eqn:DbySeis})). Then the bound is obtained by using Theorem~\ref{MainTheorem}.

If a bound is found using the background flow method, it can be also found with the auxiliary functional method with $Q[\bu,\bu]=\alpha \norm{\bu}^2/2$ and 
\begin{equation}\label{eqn:Riesz_Markov_Kakutani}
L[\ba]=-\alpha\innerproduct{\ba}{\bU}.
\end{equation}
If a bound is found using the auxiliary functional method with a quadratic auxiliary functional with  $Q[\bu,\bu]=\alpha \norm{\bu}^2/2$, it can be also found using the background flow method with $\alpha \bU$ satisfying \eq{eqn:Riesz_Markov_Kakutani}.   
This is always possible because   for a given $L$ the corresponding $\alpha\bU$ exists according to  
 the Riesz-Markov-Kakutani representation theorem, even though the background flow can turn out to be a generalized function. 

The background flow method was extended to a large variety of problems. The general conclusion that the background flow method is equivalent to the auxiliary functional method with a quadratic auxiliary functional holds in all those cases.
 \Rev{Notably, if  the nonlinear part of the governing equations conserves more than one quadratic functional, the quadratic part of the auxiliary functional is usually taken to be a sum of the conserved functionals weighted with corresponding balancing parameters.}   An up-to-date description of the background flow method is given in~\cite{fantuzzi2021background}.

\subsection{Direct method}\label{seq:direct}
The recently proposed direct method~\cite{Seis:JFM2015} is not formulated in a general form. Instead, a number of examples are given. 
The definition of the direct method given below was inferred from these examples and from~\cite{otto2011rayleigh} but is more general.
\Rev{
 It is not immediately obvious how our definition is related to that in~\cite{Seis:JFM2015,otto2011rayleigh}, but all examples in~\cite{Seis:JFM2015,otto2011rayleigh} do fit our definition, as exemplified by the sample in Section~\ref{seq:PCF} below.
}

Let $\bxi=(\xi_1,\dots,\xi_k)$ denote a vector in $\bR^k$, and introduce the notation $\bD[\ba]=(D_1[\ba],\dots,D_k[\ba])$ and  $(F[\ba],D_1[\ba],\dots,D_k[\ba])=(F[\ba],\bD[\ba])$ implying also that  $\bD[\ba]$ is an $\bR^k$-vector-valued functional of $\ba$.
\Rev{
Let $P$ be
 a set in $\bR^{k+1}$ consisting of all possible values of $(F[\ba],\bD[\ba])$, that is
$ P=\{(\zeta,\bxi)\  |\ \exists\, \ba \in A \text{ such that }(\zeta,\bxi)=(F[\ba],\bD[\ba])\}$.

\begin{lemma}\label{LemmaDMnew} 
Let 
functions $\seis_j(\zeta,\bxi),$ $j=1,\dots,n$, and a constant $B$ be such that
\begin{equation}\label{eqn:scondition1new}
\seis_j(\zeta,\bxi)\ge0    \quad\forall\,(\zeta,\bxi) \in \overline{\text{\emph{Conv}}}(P)
\end{equation}
 where $\overline{\text{\emph{Conv}}}(P)$ is the closed convex hull of $P$,
and 
\begin{equation}\label{eqn:scondition2new}
\seis_j(\zeta,0)<0 \quad\forall\,\zeta > B,j.
\end{equation}
Then $\meanT{F[\bvv(t)]}\le B$ for any $\bvv(t)\in A$ such that $\meanT{\bD[\bvv(t)]}={\bm 0}$.
\end{lemma}
}
Important note: in~\eq{eqn:scondition1new} $\bvv(t)$ is not constrained to be a solution of~\eq{eqn:dyn_sys}, and $\bD[.]$ is not constrained to be a Lie derivative of any functional.

\myproof{Since time average can be interpreted as a convex combination,\strut

\noindent 
$(\meanT{F[\bvv(t)]},\meanT{\bD[\bvv(t)]})\in  \overline{\text{{Conv}}}(P)$. The proof is then evident (by contradiction).}

\commentout{
\begin{lemma}\label{LemmaDMold} 
Let a vector-valued functional $\bD[\ba]$,  
functions $\seis_j(\zeta,\bxi),$ $j=1,\dots,n$, and a constant $B$ be such that
\begin{equation}\label{eqn:scondition1old}
\seis_j(\meanT{F[\ba(t)]},\meanT{\bD[\ba(t)]})\ge0    \quad\forall\,\ba(t),j
\end{equation}
and 
\begin{equation}\label{eqn:scondition2old}
\seis_j(\zeta,0)<0 \quad\forall\,\zeta > B,j.
\end{equation}
Then $\meanT{F[\ba(t)]}\le B$ for any $\ba(t)$ such that $\meanT{\bD[\ba(t)]}=0$.
\end{lemma}
Important note: in~\eq{eqn:scondition1} $\ba(t)$ is not constrained to be a solution of~\eq{eqn:dyn_sys}, but can be assumed to belong to the admissible set $A$ as in~\eq{eqn:constraint}. 

\myproof{Evident (by contradiction).} 
}

We define the direct method as a method consisting in finding suitable functionals $D_i[\ba]$ of the form $D_i[\ba]=\innerproduct{\bff}{\frac{\delta V_i}{\delta\ba}}$,  functions $\seis_j(\zeta,\bxi)$ and a quantity $B$ satisfying the condition of Lemma~\ref{LemmaDMnew}. 
 Since  $\meanT{\bD[\ba(t)]}={\bm 0}$ for any solution of~(\ref{eqn:dyn_sys}) by Lemma~\ref{Lemma1}, Lemma~\ref{LemmaDMnew} then gives the bound.

\Rev{ 
Similar to (\ref{eqn:optimisation}), the problem of finding the best bound by the direct method can be formulated as a problem of optimising $B$ over $V_i$ and $s_j$ subject to (\ref{eqn:scondition1new},\ref{eqn:scondition2new}), but depending on the class in which $s_j$ are sought this problem might be non-convex.
}

\Rev{
\commentout{
All the examples of the direct method implicitly use Lemma~\ref{Lemma1},  with the statement of the lemma proved anew for each particular case without explicitly introducing the auxiliary functional. The first step of the direct method leads to a set of statements of the form  $\meanT{D_i[\bu(t)]}=0$ for any solution $\bu(t)$. This is the only step where the governing equations are used. The  functionals $D_i$ might later be combined into a single one.  
The second step of the direct method is the derivation of an inequality or inequalities involving $\meanT{F[\bvv(t)]}$ and $\meanT{D_i[\bvv(t)]}$ valid for any $\bvv(t)\in A$ for which the averages exist. Crucially, functions $\bvv(t)$ are not necessarily solutions of the governing equations. A distinguishing feature of the direct method is that these inequalities might be nonlinear in  $\meanT{F[\bvv(t)]}$ and $\meanT{D_i[\bvv(t)]}$, unlike the auxiliary function and background flow methods. 
The third step is the derivation of the bound for  $\meanT{F[\bu(t)]}$ from the inequalities obtained combined with the conditions that $\meanT{D_i[\bu(t)]}=0$ that were proved in step one to be valid for any solution $\bu(t)$.

}
Examples of the direct method~\cite{Seis:JFM2015,otto2011rayleigh} do not use Lemma~\ref{LemmaDMnew} explicitly. They give, however, important highlights of the ways and motivations on how  $D_i$ and $s_j$ can be selected and the corresponding inequalities can be proved. The definition of the direct method could be narrowed to the specific techniques used in~\cite{Seis:JFM2015,otto2011rayleigh}. Instead, our definition abstracts from these details in favour of simplicity and generality helping to prove the equivalence theorems below. 
     
}


\subsection{Relationship of the auxiliary functional method and the direct method}\label{seq:auxiliaryVSdirect}

\setlength{\unitlengthstored}{\unitlength}
\setlength{\unitlength}{0.6\textwidth}
\begin{figure}[h!]
\centering
\begin{picture}(1,0.556)
\put(0,0){\includegraphics[width=0.5\textwidth, trim = 250 190 120 110, clip]{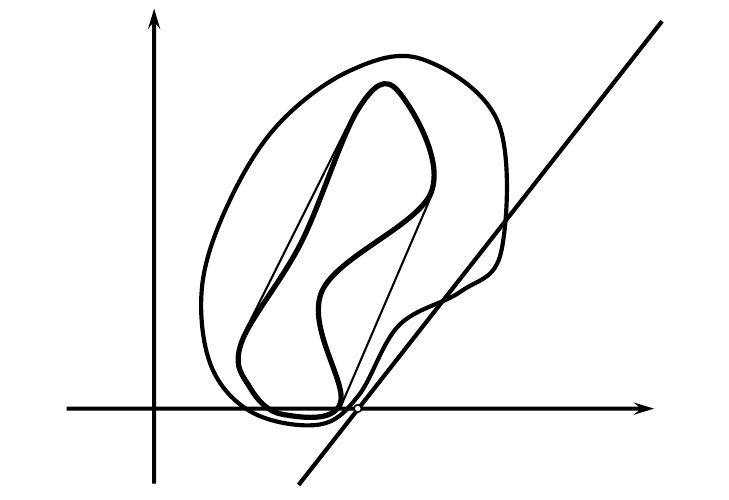}}
\put(0.1,0.5){$\xi$}
\put(0.7,0.05){$\zeta$}
\put(0.4,0.05){$B$} 
\put(0.6,0.3){\rotatebox{52}{$\zeta-B+\alpha \xi = 0$}}
\put(0.51,0.4){$S$}
\put(0.420,0.4){$P$}
\end{picture}
\caption{\label{fig:figure} \Rev{
Set $P$ is the set of all $(F[\bu),D[\bu])$. $\OTP$ consists of $(\meanT{F[\bvv(t)]},\meanT{D[\bvv(t)]})$ for all $\bvv(t)$. For solutions, $\meanT{D[\bu(t)]}=0$. Hence solutions points have $\xi=0$ and are in $\OTP$. The direct method gives the set $S\supseteq \OTP$ and the value $B\ge \zeta\, \forall\,  (\zeta,0)\in S$.  The line demarcates the constraint of the auxiliary function method.
}
}
\end{figure}
\setlength{\unitlength}{\unitlengthstored}

 \begin{theorem}\label{TheoremAFMtoDM}
 If a certain bound is found by the auxiliary functional method, the same bound can be obtained by the direct method.
\end{theorem} 
\myproof{ 
If $B$ is the bound obtained by the auxiliary functional method then,  \Rev{
with $k=n=1$ and  $\seis_1(\zeta,\xi_1)=B-\zeta-\xi_1$, (\ref{eqn:constraint})  proves the inequalities~(\ref{eqn:scondition1new}) and (\ref{eqn:scondition2new}) of the direct method, and the same bound follows from Lemma~\ref{LemmaDMnew}.
}
  }

Understanding the proofs of the following converse theorems can be simplified by a geometric interpretation (see figure~\ref{fig:figure}). 
Consider a set 
$\displaystyle
S=\{(\zeta,\bxi)\ |\ \seis_j(\zeta,\bxi)\ge0 \ \forall j\} \subset \bR^{k+1}
$, that is the set of points satisfying \eq{eqn:scondition1new}.
\commentout{ =======================
,
and a set
$
\displaystyle
 \meanT{P}=\{(\zeta,\bxi)\  |\ \exists\, \ba(t) \text{ such that }(\zeta,\bxi)=(\meanT{F[\ba(t)]},\meanT{\bD[\ba(t)]})\}
$,
that is a set in $\bR^{k+1}$ consisting of all possible values of $(\meanT{F[\ba(t)]},\meanT{\bD[\ba(t)]}).$
}
Then \eq{eqn:scondition1new} implies $\OTP \subseteq S$, and the direct method involves finding the point $(B,\bm 0)$  to the right from $S$ (or at the ``right border" of $S$).
 \begin{theorem}\label{TheoremDMtoAFM}
If a bound $B$ is found by the direct method, the same bound can be obtained by the auxiliary  functional method provided that the functions $\seis_j$ satisfy the condition  
\begin{equation}\label{eqn:AMequivalence}
 \exists\, \alpha_i \in \bR: \zeta-B+\alpha_i\xi_i\le0\   \forall\, \zeta,\bxi: \seis_j(\zeta,\bxi)\ge0 \ \forall j.
\end{equation}
(This can also be written as $ \exists\, \alpha_i \in \bR : \zeta-B+\alpha_i\xi_i\le0\   \forall\, (\zeta,\bxi) \in S$.)

\end{theorem}
\myproof{ 
From (\ref{eqn:scondition1new}) it follows that the same inequalities are valid for arbitrary time-independent $\ba$:
$\displaystyle
\seis_j(F[\ba],\bD[\ba])\ge0 \quad\forall j, \ba
$.
From (\ref{eqn:AMequivalence}) it follows that
\begin{equation}\label{eqn:alphaDcondition}
 \exists\, \alpha_i \in \bR: F[\ba]-B+\alpha_iD_i[\ba]\le0\qquad \forall \ba
\end{equation}
and hence  $\meanT{F[\ba(t)]}\le B$ by Theorem~\ref{MainTheorem}. }

\begin{remark}\label{remark:geometric}
Condition (\ref{eqn:AMequivalence}) means that there is a plane $F[\ba]-B+\alpha_iD_i[\ba]=0$ such that no point in $S$ is located to the right of that plane, that is in the side with greater values of $\zeta$  (see figure~\ref{fig:figure}). 
From this interpretation it follows (for example from the supporting or separating hyperplane theorems)  that for (\ref{eqn:AMequivalence}) to be satisfied it is sufficient (but not necessary) for the set  $S$
 to be convex, for which in turn it is sufficient for all the functions $\seis_j(\zeta,\bxi)$ to be concave. The separating  hyperplane theorem applies when $(B,\bm0)\notin S$.  It states that there exists $\beta$ and $(\gamma_1,\dots,\gamma_k)$ such that $\beta\zeta+\gamma_i\xi_i\le \beta B$ for all $(\zeta,\bm\xi)\in S$.
  Taking $\bm\xi=0$ in this and (\ref{eqn:AMequivalence}) gives   $\beta\zeta\le \beta B$ and $\zeta\le B$ for all $(\zeta,\bm 0)\in S$. This implies $\beta>0$ in all but the trivial case when the only point in $S$ such that $\xi=0$ is $(B,\bm 0)$. Then $\alpha_i=\gamma_i/\beta$ satisfy (\ref{eqn:AMequivalence}). The supporting hyperplane theorem can be used when $(B,\bm 0)$ is on the border of $S$. 
\end{remark}
\begin{remark}
The auxiliary functional corresponding to~(\ref{eqn:alphaDcondition}) is, of course, $\alpha_iV_i[\ba].$ Hence, for convex $S$ Theorem~\ref{TheoremDMtoAFM} is constructive. 
\end{remark}

If the set $S$ is not convex, it is not guaranteed that the conditions of Theorem~\ref{TheoremDMtoAFM} can be satisfied. However, the following non-constructive theorem is valid:
 \begin{theorem}\label{FullEquivalenceTheorem}
If a bound was obtained by the direct method with a certain set of $D_i[\ba]=\innerproduct{\bff}{\frac{\delta V_i}{\delta\ba}}$ then there exists $\alpha_i$ such that the same bound can be obtained with the auxiliary functional method with the auxiliary  functional $V=\alpha_iV_i$.
\end{theorem}
\myproof{
Coincides with the proof of Theorem~\ref{TheoremDMtoAFM} with $S$ replaced with $\OTP\subseteq S$.
}

\begin{remark}
The difference between this result and Theorem~\ref{TheoremDMtoAFM} is that while the set $S$ is known from the result obtained by the direct method, thus giving a method of calculating $\alpha_i,$  $\text{Conv}(P)$ is not known. However, even when $S$ is not convex, the proof of a bound by the direct method still provides $V_i$.
\end{remark}  

Essentially,  the auxiliary functional method and the direct method are equivalent, as follows from 
Theorems~\ref{TheoremAFMtoDM}, \ref{TheoremDMtoAFM} and~\ref{FullEquivalenceTheorem}.

\section{Plane Couette flow}\label{seq:PCF}

To the best of  our knowledge, so far all existing implementations of the direct method involved dynamical systems with a quadratic nonlinearity and (implicitly) used only linear and quadratic auxiliary functionals in the analysis. Therefore, all the results so far obtained by the direct method can also be obtained by the background flow method. We will now illustrate this observation with an analysis of the plane Couette flow.

\subsection{Problem formulation}

We will insignificantly generalize the reasoning in \cite{Seis:JFM2015} of obtaining the bound for energy dissipation rate in the plane Couette flow, staying as close to \cite{Seis:JFM2015} as possible for our purposes. We will consider the solutions of the Navier-Stokes equations 
\begin{equation}\label{eqn:NSE}
\pd{\bu}t+\bu\cdot\nabla\bu=-\nabla p + \nabla^2 \bu
\end{equation}
and the continuity equation 
\begin{equation}\label{eqn:continuity}
\nabla\cdot\bu=0
\end{equation}
in the domain $0\le x \le L,$ $0\le y \le L,$ $0\le z\le 1,$ where $x,$ $y,$ and $z$ are Cartesian coordinates, and $\bu=(u,v,w)$ is the velocity field.  
The lower wall is located at $z=0,$ where the no-slip condition $\bu=0$ is imposed. The upper wall at $z=1$ is moving at a constant speed $\Rey$, so that the boundary condition is $u=\Rey,$ $v=w=0.$ The required solution is periodic in $x$ and $y$ with the periods $L.$

We will distinguish the horizontal average defined as
\[
\meanH{f(t,x,y,z)}=\frac1{L^2}\iint_{x,y=0}^{x,y=L} f(t,x,y,z)\,dx\,dy
\]
{and the time and horizontal average: 
}  
$
\meanHT{f(t,x,y,z)}
$.
The spatial average of the energy dissipation rate is
$
\dissiIns[\bu]=\zint{\meanH{|\nabla\bu|^2}}$,
and the time-averaged dissipation rate is
$
\dissiAve=\meanT{\dissiIns[\bu(t)]}.
$
The problem consists in deriving an upper bound for $\dissiAve.$


\Rev{ 

\subsection{\Rev{Lie derivative of the auxiliary functional for plane Couette flow}}


We will loosely follow \cite{Seis:JFM2015}, but without time averaging. 
Multiplying (\ref{eqn:NSE}) by $\bu,$ taking horizontal average, integrating by parts with the periodic boundary conditions used, integrating with respect to $z$ and using the boundary conditions gives the energy equation
\begin{equation}\label{eqn:S3.5}
\der{\ }t\zint{\frac{\meanH{\bu^2}}2}=\Rey{\left.\pd{\meanH{u}}z\right|_{z=1}}-\dissiIns[\bu],
\end{equation}
which corresponds to\ (3.5) in\  \cite{Seis:JFM2015} in the sense that it becomes \ (3.5) in\  \cite{Seis:JFM2015} after time averaging. 

Taking the horizontal average of the $x$-component of  (\ref{eqn:NSE}) gives\footnote{This is the well-known Reynolds-averaged Navier-Stokes equation for a unidirectional mean flow depending on one coordinate.} 
\begin{equation}\label{eqn:S3.6}
\pd{\ }t\meanH u=-\pd{\meanH{uv}}z+\pdd{\meanH u}z \quad\text{for all }z,
\end{equation}
which corresponds to\ (3.6) in\  \cite{Seis:JFM2015}. 
Integrating~(\ref{eqn:S3.6}) with respect to $z$ and using the boundary 
conditions gives
\begin{equation}\label{eqn:star}
\pd{\ }t\int_1^z{\left.{\meanH u}\right|_{z=\zeta}}\,d\zeta=-\meanH{uv}+\pd{\meanH u}z-\left.\pd{\meanH u}z\right|_{z=1}\quad\text{for all }z.
\end{equation}
Adding~(\ref{eqn:star}) multiplied by $\Rey$ to (\ref{eqn:S3.5}) gives
%
\begin{equation}
\label{eqn:S3.7}
\pd{\ }t\left(\zint{{\frac{\meanH{\bu^2}}2}}-\Rey\int_z^1{\meanH{u(t,x,y,\zeta)}\,d\zeta}\right)= \\
-\dissiIns[\bu]+\Rey\meanH{\pd uz-uw} \quad\text{for all }z,
\end{equation}
\noindent which corresponds to\ (3.7) in\  \cite{Seis:JFM2015}.
We will now introduce the weight $\rho(z)$ such that
\[
\zint{\rho}=1,
\]
and take a weighted $z$-average of (\ref{eqn:S3.7}), arriving at
\begin{equation}
\label{eqn:S3.8}
\der{\ }t\zint{\left({\frac{\meanH{\bu^2}}2}-\Rey\rho(z)\int_z^1{\meanH{u(t,x,y,\zeta)}d\zeta}\right)\!}= \\
-\dissiIns[\bu]+\Rey\zint{\rho(z)\meanH{\pd uz-uw}\!},
\end{equation}
\noindent which corresponds to\ (3.8) in\  \cite{Seis:JFM2015} if $\rho(z)$ is taken to be
\begin{equation}\label{eqn:rhoS}
 \rho(z)=
  \begin{cases}
  1/l,& 0\le z\le l \\
 0,& l < z\le 1.
  \end{cases}
 \end{equation}
Hence, the auxiliary functional corresponding to the analysis of~\cite{Seis:JFM2015} is
\begin{equation}\label{eqn:VD}
 V_D[\bvv]=\zint{\left({\frac{\meanH{\bvv^2}}2}-\Rey\rho(z)\int_z^1{\meanH{\tu(t,x,y,\zeta)}\,d\zeta}\right)},
\end{equation}
 and, as follows from  (\ref{eqn:S3.8}), its Lie derivative with respect to
 (\ref{eqn:NSE},\ref{eqn:continuity}) is
 \begin{equation}\label{eqn:DD}
 D_D[\bvv]=-\dissiIns[\bvv]+\Rey\zint{\rho(z)\meanH{\pd {\tu}z-\tu\tw}}.
 \end{equation}

We added tildes to highlight that while $\bu=\bu(t)$ in (\ref{eqn:S3.8}) is the solution of (\ref{eqn:NSE},\ref{eqn:continuity}), $\bvv$ is any function in $A$, including functions independent of time.
Functional $V_D[\bvv]$ can be represented in a form accepted in the background flow method and with a multiplicative coefficient as 
\begin{equation}\label{eqn:VB}
 V_B[\bvv]=\alpha E[\bvv-\bU]+\text{const}=\alpha V_D[\bvv] 
 \end{equation}
 with $\bU=(U(z),0,0)$ and 
\begin{equation}\label{eqn:Urho}
 \der Uz=\rho(z)\Rey,\quad U(0)=0.
 \end{equation}
 This can be verified by substituting~(\ref{eqn:Urho}) into~(\ref{eqn:VD}) and integrating by parts.  For the weight~(\ref{eqn:rhoS}) used in\  \cite{Seis:JFM2015} the corresponding background flow is
 \[
 U(z)=
  \begin{cases}
  \Rey z/l,& 0\le z\le l, \\
 \Rey,& l < z\le 1.
  \end{cases}
  \]

\subsection{Reduction to our definition of the direct method}\label{seq:reductiontoourdef}
In~\cite{Seis:JFM2015} two estimates for the time averages
\begin{equation}\label{eqn:Idef}
I=\Rey\zint{\rho(z)\meanHT{\pd {\tu}z-\tu\tw}}=\meanT{D_D[\bvv(t)]}+\meanT{\epsilon[\bvv(t)]}
\end{equation}
 of the integral in the right-hand side of (\ref{eqn:DD}) are obtained. The key observation is that the derivation of these estimates does not use the governing equations. Hence, these estimates  are valid for any function $\bvv(t)$ and not only for those $\bu(t)$ that satisfy~(\ref{eqn:NSE}-\ref{eqn:continuity}).
The first estimate is
\begin{equation}\label{eqn:S1}
I\le C\Rey\left(\frac1{l^{1/2}}\meanT{\epsilon[\bvv(t)]}^{1/2}+l\meanT{\epsilon[\bvv(t)]}\right)
\end{equation}
corresponding to (3.12) in \cite{Seis:JFM2015}\footnote{With a constant $C$ added to switch from $\lesssim$ to $\le$ notation more convenient in our context.}. We also take $l=1/(3C\Rey)$, which is the optimal value.
We choose $s$ as 
\[
s(\zeta, \xi)=-\xi -\frac23 \zeta+ \sqrt{3C^3\Rey^3}\zeta^{1/2}.
\]
From (\ref{eqn:Idef}) and (\ref{eqn:S1}) it follows that 
\[
s\left(\meanT{\epsilon[\bvv(t)]},\meanT{D_D[\bvv(t)]}\right) \ge 0 \quad \forall \bvv(t).
\]
This satisfies the first condition (\ref{eqn:scondition1new}) of Lemma~\ref{LemmaDMnew}. 
Note that $s(\zeta, \xi)$ is concave.
Simple algebraic analysis gives that
\[
 s(\zeta, \xi) \ge 0 \Rightarrow    \zeta - \frac{27}4C^3\Rey^3 +3 \xi \le 0, 
\]
which is the condition (\ref{eqn:AMequivalence}) of Theorem~\ref{TheoremDMtoAFM} with $\alpha=3$.  Hence, by this theorem, the auxiliary function method with the auxiliary functional equal to $
3 V_D$ proves the bound 
\[
\dissiAve \le  \frac{27}4C^3\Rey^3.
\]

On the other hand, following~\cite{Seis:JFM2015} the same bound can be obtained directly, by noticing that 
if $\bvv(t)=\bu(t)$, that is it is the solution of~(\ref{eqn:NSE}-\ref{eqn:continuity}), then $\meanT{D_D[\bvv(t)]}=0$ and, therefore,
$s(\dissiAve,0)=-\frac23 \dissiAve+ \sqrt{3C^3\Rey^3}\dissiAve^{1/2} \ge 0.$

} 

\section{Discussion with a look beyond quadratic functionals}\label{sec:discussion}
\subsection{Advantages of the three methods}

As we have shown, bounds obtained with the auxiliary functional method can also be obtained with the direct method and vice versa. Moreover, as it has already been known~\cite{CGHP:RoySoc2014}, if the auxiliary functional is quadratic, the auxiliary function method effectively coincides with the background flow method. (It can be verified that the auxiliary functionals corresponding to the direct method application to the Rayleigh-B\'enard convection in the infinite-Prandtl-number limit~\cite{otto2011rayleigh} are quadratic and in some cases even linear).  In~\cite{nobili2017limitations} it is stated that the best bound obtainable with the background flow method for the same problem is not as good as the bound obtained in~\cite{otto2011rayleigh} using the direct method.  This, however, does not contradict our results, because due to the application of the maximum principle the admissible set $A$ in~\cite{otto2011rayleigh} is a proper subset of  the admissible set $A$ implied by the definition of the background flow method adopted in~\cite{nobili2017limitations}.

The direct method has an advantage of eliminating by time averaging the time derivatives of the auxiliary functional early in the analysis, reducing the amount of analytic work required. This can be quite substantial, as, for example,  in~\cite{otto2011rayleigh}.

 The main advantage of the auxiliary functional method is its conceptual simplicity, which allows to move on to more difficult problems and more complicated auxiliary functionals. For example, reformulating the background flow method in terms of the auxiliary functional method led to a definite progress in developing the background flow method itself~\cite{fantuzzi2021background}.

In principle both the auxiliary functional method and the direct method can give better bounds than the background flow method when non-quadratic functionals are used.  A specific example of this is the results for the Kuramoto-Sivashinski equation~\cite{Goluskin_2019}.

The background flow method, being equivalent to the auxiliary functional method with a quadratic auxiliary functional, has an advantage when the functional to be bounded is also quadratic,  as for example it is the case when the bound for the energy dissipation rate is sought for. In such cases the constraint in the bound optimization problem becomes quadratic, which is much  easier to treat both numerically and analytically. So far, when partial differential equations are concerned, the background flow method (and its equivalent quadratic auxiliary functional and direct method) remains the only general method yielding to fully analytic treatment. It is attractive to utilize the experience accumulated with background flow method while making progress with a more general auxiliary functional method. This deserves a more detailed discussion.

\subsection{Non-quadratic functionals}

As we explained in the Introduction, when both the upper $B_U$ and lower $B_L$ bounds for a time average are sought for, 
having a trade-off between the margin  $B_U -B_L$ and the computational cost is desirable.  Quadratic auxiliary functionals, however, do not provide such a trade-off for two reasons.

First, as mentioned in \cite{FantuzziEtAl:JADM2016} (see (7.3) in that paper) in application to finite-dimensional systems, if a bound is sharp then $\innerproduct{\bff[\bu]}{\varder{V[\bu]}{\bu}}=\meanT{F} - F[\bu]$ along the extremal periodic solution. This is a very strong constraint that a  quadratic $V$ often cannot satisfy even approximately.  In contrast, the auxiliary function method can give arbitrarily sharp  bounds on time averages~\cite{tobasco2018optimal,rosa2020optimal}.  

Second, if the system has two solutions with different time averages then the bounds obtained with the methods discussed above have to be valid for both solutions and, hence, the difference $B_U-B_L$ cannot be made arbitrarily small. In practice one of the solutions is often unstable, as it is the case with laminar and turbulent flows in fluid dynamics. Then the ever-present small noise makes the unstable solution irrelevant. To benefit from this one needs to extend the auxiliary functional method to the case of systems with noise. Thanks to the conceptual simplicity of the auxiliary functional method this can be easily done~\cite{CGHP:RoySoc2014,FantuzziEtAl:JADM2016} at least for finite-dimensional systems. It turns out, however, that if the auxiliary functional is quadratic then the resulting bound for the system with noise tends to the bound for the non-stochastic system as the noise intensity tends to zero. This is not the case if the auxiliary functionals are not limited to quadratic functionals. (The readers interested in systems with noise should also see~\cite{kuntz2016bounding,korda2021convex}, which will also lead to a large volume of related works.)  

The advantages of using higher-than-quadratic auxiliary polynomials were  demonstrated on a number of specific examples including finite-dimensional and infinite-dimensional systems in \cite{CGHP:RoySoc2014,FantuzziEtAl:JADM2016,goluskin2018bounding,HuangEtAl:RSPA2015,GC:PhysD2012,Goluskin_2019}. 
A systematic way of constructing higher-degree polynomial auxiliary functionals for infinite-dimensional systems was proposed in \cite{GC:PhysD2012} and implemented in \cite{HuangEtAl:RSPA2015} in the context of Lyapunov stability of fluid flows. In~\cite{CGHP:RoySoc2014} it was extended to bounds of time averages and applied in~\cite{Goluskin_2019} to bounds on mean energy in the {K}uramoto-{S}ivashinsky equation. This method relies on polynomial semi-algebraic optimization and is computer-assisted, and thus it is less suitable for analytic investigations. In the following sections we propose three other approaches to constructing non-quadratic, and generally non-polynomial, auxiliary functionals, which retain the main features of the background flow method and thus might inherit some of its advantages.

\subsection{Extending the background flow method to varying balance parameter}\label{sec:balanceextention}
Let us revisit the form of the auxiliary functional~(\ref{eqn:Vbackground}) in the standard background flow method, which we rewrite in a slightly different but equivalent form: 
\begin{equation}\label{eqn:Vbackground2}
V[\bu]=\alpha \norm{\bu}^2/2-\innerproduct{\bu}{\bU}.
\end{equation}
This is obtained by dropping the irrelevant term independent of $\bu$ and incorporating the balance parameter into the background flow $\bU$ in the second term.
The optimization constraint (\ref{eqn:constraint}) then has the form
\begin{equation}\label{eqn:constraint2}
F[\bu]-B+\alpha\innerproduct{\bl(\bu)+\br}{\bu}-\innerproduct{\bff(\bu)}{\bU}\le 0 \qquad \forall \bu \in A.
 \end{equation}
 Conveniently, if $F[\bu]$ is a quadratic functional then (\ref{eqn:constraint2}) is quadratic, too.

Let us now take the auxiliary functional in a more general form
\begin{equation}\label{eqn:VbackgroundNonlinear}
V[\bu]=a(\norm{\bu}^2/2)-\innerproduct{\bu}{\bU},
\end{equation}
where $a(.)$ is an arbitrary differentiable function. Then the optimization constraint becomes
\begin{equation}\label{eqn:constraintNonlinear}
F[\bu]-B+a'(\norm{\bu}^2/2)\innerproduct{\bl(\bu)+\br}{\bu}-\innerproduct{\bff(\bu)}{\bU}\le 0 \qquad \forall \bu,
 \end{equation}
which coincides with (\ref{eqn:constraint2}) with $\alpha= a'(\norm{\bu}^2/2).$ In other words, the balance parameter in the background flow method need not be a constant. Instead it can be an arbitrary function of energy. Therefore, first $B$ can be minimised over $\alpha_q$ and $\bU$  subject to a weaker constraint
\begin{equation}\label{eqn:constraint3}
F[\bu]-B_q+\alpha_q\innerproduct{\bl(\bu)+\br}{\bu}-\innerproduct{\bff(\bu)}{\bU}\le 0 \qquad \forall \bu\in A  \text{ such that } \norm{\bu}^2=q.
 \end{equation}  
This results in the optimisation of the bound being done for a smaller admissible set $ \{\bu\, |  \norm{\bu}^2=q\}\cap A$ for each $q$. However, the constraint itself remains the same as in the standard background flow method. Then the bound is $B=\sup_q B_q,$ which can be better than the bound obtained with the stronger constraint~(\ref{eqn:constraint2}). 

\subsection{Extending the background flow method to varying background flow}
Obviously, Lemma~\ref{Lemma1} can be extended to continuous and piecewise-differentiable auxiliary functionals. (To be precise, the bound obtained with the auxiliary function method is valid for all bounded solutions $\bu(t)$ on which  $V[\ba(t)]$ is a piecewise-differentiable function of time. If the auxiliary functional does not ensure this on certain trajectories, the bound might be invalid for these trajectories, but it will still be valid for the trajectories satisfying this condition.) That allows to generalise \eq{eqn:Vbackground} to 
\begin{equation}\label{eqn:V2backgrounds}
V[\ba]=\begin{cases}\alpha{\norm{\ba-\bU_1}^2}/2, &\norm{\ba-\bU_1}<\norm{\ba-\bU_2},\\
                                \alpha{\norm{\ba-\bU_2}^2}/2, &\text{otherwise.}
           \end{cases}
\end{equation}
In geometric interpretation, this auxiliary functional is proportional to the squared distance to the nearest of the two background flows. As in \S\ref{sec:discussion}\ref{sec:balanceextention}, the constraint keeps the same form as in the original background flow method, but there are two admissible sets corresponding to $\bU_1$ and $\bU_2$. This idea can be generalised to any, including infinite, number of background flows.

Remarkably, long before the background flow method has been interpreted in terms of auxiliary functionals, Goodman~\cite{goodman1994stability} used an infinite set of background flows in order to extend the bound for the energy of the odd periodic solutions of  the Kuramoto–Sivashinsky equation to all periodic solutions. In general, this amounts to choosing a set~$\,\mathcal{U}$ and taking the time-dependent background flow $\bU\in\mathcal{U}$ to be the element nearest to $\ba$, that is
\begin{equation}\label{eqn:GoodmanU}
\bU=\argmin_{\bW \in \mathcal{U}} \norm{\bW-\ba}.
\end{equation}
 In this case the Lie derivative of the auxiliary functional is 
\[
\der{\ }t\left[\frac{\alpha}2\norm{\ba-\bU}^2\right]=\alpha\innerproduct{\ba-\bU}{\bff({\ba})}-\alpha\innerproduct{\ba-\bU}{\der{\bU}t}=\alpha\innerproduct{\ba-\bU}{\bff({\ba})}.
\]      
 The second term  in the middle is zero because $d{\bU}/dt$ is either zero or it is perpendicular to $\ba-\bU$, so that $\innerproduct{\ba-\bU}{d\bU/dt}=0$ \Rev{when $d{\bU}/dt$ exists}. A specific example is in~\cite{goodman1994stability}. The geometric interpretation is that the vector $\ba-\bU$ connecting the point $\ba$ with the nearest point in $\mathcal{U}$ is perpendicular to the boundary of $\mathcal{U}$. When $\ba$ moves, the nearest point $\bU\in\mathcal{U}$ either moves along the smooth boundary, jumps without changing $V$ as in \eq{eqn:V2backgrounds}, or stays at a corner. If $\bU$ happens to be at a corner of the boundary it can remain constant, but then $d\bU/dt=0$.

The auxiliary functionals corresponding to \eq{eqn:V2backgrounds} and \eq{eqn:GoodmanU} depend on $\ba$ non-polynomially, but the corresponding constraints 
have the same or very similar form to the constraint of the background flow method.

 \subsection{Helicity as a part of an auxiliary functional}\label{sec:helicity}

There are only two known nontrivial conservative quantities in three-dimensional incompressible inviscid 
flows: kinetic energy and helicity. Helicity per unit volume is defined as $h=\bu\cdot\bomega$,
where $\bomega=\rot \bu$ is vorticity. It can be shown that in inviscid flows it satisfies the equation that can be written in a conservative form as
$\partial h/\partial t=-\nabla\cdot\left(\bu h -\bomega(p-|\bu|^2/2) \right)$,
where $p$ is the pressure made non-dimensional with the density and the characteristic velocity~\cite{moffatt1992helicity}. 
Helicity of a fluid flow in a domain $\Sigma$ is $H=\int_\Sigma h\,d\Sigma$,
and therefore
\begin{equation}\label{eqn:dHdt}
\der Ht=-\int_{\partial \Sigma} u_nh-\omega_n (p-|\bu|^2/2)\,d\sigma,
\end{equation}
where the subscript $n$ denotes the normal components of the vectors at the boundary of the domain. The right-hand side of~\eq{eqn:dHdt} can be zero for certain boundary conditions, including for example periodic boundary conditions. However, most interesting boundary conditions usually involve solid walls. While in inviscid flows $u_n=0$ on a solid wall, neither $\omega_n$ nor $p-|\bu|^2/2$ can be guaranteed to be zero on the wall, and in such cases helicity in the domain of an inviscid flow is not conserved. This appears to leave the kinetic energy as the only possible basis for applying auxiliary function methods with a quadratic functional to fluid flows. Fortunately, in viscous flows on solid wall with the no-slip conditions $\omega_n = 0$~\cite{moffatt1969degree}. Therefore, one can attempt to extend the background flow method by taking 
$Q[\bu,\bu]=\alpha {\norm{\bu}^2}/2+\beta H[\bu]$,
which will satisfy the neccesary condition \eq{eqn:dEdt} in viscous flows with no-slip condition, even though not for inviscid flows.
However, such an attempt cannot succeed, because $H[\bu]$ and, moreover, $\alpha {\norm{\bu}^2}/2+\beta H[\bu]$ are unbounded, while according to Proposition A.2 in~\cite{Goluskin_2019} auxiliary functional proving an upper bound has itself to be bounded from below,  
under the assumptions that are satisfied for typical viscous flows.

The idea of \S\ref{sec:discussion}\ref{sec:balanceextention} might help to overcome this difficulty. Namely, one can seek the auxiliary functional in the form
\begin{equation}
V=\alpha \left(\frac{\norm{\bu}^2}2\right)\frac{\norm{\bu}^2}2+\beta (H[\bu])H[\bu]+\dots,
\end{equation}
that is with the balancing parameters being functions of the corresponding arguments, and specify $\beta(.)$ such as to make $V$ bounded as required. 

The approaches proposed in this section might allow to obtain better bounds while utilizing the experience accumulated in the studies based on the standard background flow method.


\section{Conclusion}

It has already been known that the background flow method of finding bounds for time averages is a special case of the auxiliary functional method when the auxiliary functional is quadratic. In the present work it has been shown that the auxiliary functional method and the direct method are equivalent. The constructive proof is given of the theorem stating that a bound obtained with the auxiliary functional method can also be obtained by the direct method. The converse theorem is also constructive under a mild assumption. In the general case when this assumption is not satisfied, the proof of the bound by the direct method allows to determine, constructively, a set of functionals such that their linear combination is an auxiliary functional proving the same bound. However, the proof of the existence of such coefficients given in the present paper is not constructive.  
These general conclusions are illustrated with an analysis of the plain Couette flow.

While the background flow method is only a special case of the auxiliary functional method, it is much easier to use in analytic studies. At the same time, further progress in the area of obtaining bounds for time-averaged characteristics of dynamical systems is likely to be associated with non-quadratic functionals. Three approaches  combining the advantages of these two methods are proposed. The first approach involves making the balancing parameter a function of the energy. The second approach consists in making the background flow to be the element of a given set of flows that minimises the instantaneous value of the norm of the difference between the actual flow and the background flow. The third approach involves adding a function of helicity to the auxiliary functional.  
\vskip6pt

\enlargethispage{20pt}




\centerline{--}


The author declares that he has no competing interests.


\centerline{--}

The author is grateful to Giovanni Fantuzzi and Andy Wynn for insightful comments to an early version of this paper, to David Goluskin for bringing the paper by Goodman to author's attention, and to the anonymous referees for insightful comments. 




\bibliography{Chernyshenko_bib_2021}{}
\bibliographystyle{plain}

\end{document}